**Redrawing the "Color Line": Examining Racial Segregation in Associative Networks on Twitter**

Nina Cesare[1], Hedwig Lee[1,2], Tyler McCormick, [1,2,3] Emma S. Spiro,[1,4]

[1] Department of Sociology, University of Washington
[2] Center for Studies in Demography and Ecology, University of Washington
[3] Department of Statistics, University of Washington
[4] The Information School, University of Washington

**ABSTRACT**

Online social spaces are increasingly salient contexts for associative tie formation. However, the racial composition of networks within most of these spaces has yet to be examined. In this paper, we analyze racial segregation patterns in associative networks on Twitter. Acknowledging past work on the role that social structure and agency play in influencing the racial composition of individuals' networks, we argue that Twitter blurs the influence of these forces and may invite users to generate networks that are more or less segregated than what has been observed offline, depending on use. While we expect to find some level of racial segregation within this space, this paper unpacks the extent to which we observe same-race connectedness for black and white users, assesses whether these patterns are likely generated by opportunity or by choice, and contextualizes results by comparing them with patterns of same-race connectedness observed offline.

**INTRODUCTION**

Despite policymakers' best efforts to derail a legacy of racial inequality in the United States (US), neighborhoods, schools, and other social contexts continue to be defined along racial lines. In particular, most individuals' social networks remain racially homogeneous even among cohorts born decades after the dismantling of Jim Crow and federal regulation of other discriminatory practices and policies (Moody, 2001; Quillian & Campbell, 2003). While no one causal explanation exists, a large body of sociological research examining friendship segregation across a variety of contexts concludes that both individual preferences and structural barriers help to either foster or reduce the existence of cross-race ties within associative networks (McPherson, Smith-Lovin, & Cook, 2001; Moody, 2001; Quillian & Campbell, 2003; Wimmer & Lewis, 2010).

While the macro- and micro-level factors associated with patterns of racial segregation in social networks offline are well examined, social scientists have yet to explore whether similar patterns appear within social media sites – online spaces that allow users to "(1) construct a public or semi-public profile within a bounded system, (2) articulate a list of other users with whom they share a connection, and (3) view and traverse their list of connections and those made by others within the system" (boyd & Ellison, 2007: 211).

In regards to their capacity to facilitate interaction, social media spaces have unique constraints and affordances. While some sites are "nonymous," meaning that users online self-presentation is strongly anchored in their offline social identity (Zhao, Grasmuck, & Martin, 2008), others are more anonymous and allow individuals to strategically display or conceal parts of their offline selves and selectively reach out to others with or without the desire for reciprocation. Indeed, some have proposed social media spaces are in fact unique social contexts that constitute a "habitus of the new" where social structure and individual agency are perpetually co-evolving (Papacharissi & Easton, 2013; Taylor-Smith, 2012). Overall, the unique structure of social media spaces invites consideration of whether or not patterns of segregation within these spaces are similar to those observed within offline networks.

This study focuses on the composition of network ties within social media. It first reviews literature examining the role of structure and agency in the development of friendship networks and argues that social media spaces, particularly Twitter, may blur the role of these forces in shaping network composition. Within online contexts users interact in a social space where they are actively engaged in constructing their own "structure" in the form of norms and expectations (Papacharissi & Easton, 2013; Taylor-Smith, 2012). Moreover, within some sites users may have the opportunity to connect with others outside of their immediate offline social context, thus reducing the role of structural barriers affecting network integration. Indeed, many describe social media as being "borderless" – meaning that they facilitate communication patterns that transcend geographic bounds. While it is likely that online networks exhibit some degree of segregation, it is difficult to anticipate the patterns of segregation that may be observed. Furthermore, the degree to which these connections are formed as the result of opportunity or choice necessitates unpacking.

This study examines patterns of friendship segregation within Twitter – a popular microblogging space. We elect to study Twitter due primarily to the fact that Twitter offers directed rather than reciprocal ties, which help lower the cost of connecting with others and may encourage the formation of diverse networks. To explore segregation on Twitter, we first randomly sample a set of Twitter users, obtain information on the friends and followers of these users, and extract demographic information from these profiles using Face++ facial recognition software (www.faceplusplus.com) and crowdsourced human evaluations (McCormick, Lee, Cesare, Shojaie, & Spiro, 2015). We then assess the same-race connectedness of black and white users, and unpack whether the levels of same-race connectedness we observe are likely the result of opportunity/propinquity or user choice. Finally, to contextualize these results we compare the racial composition of these networks to those reported for offline networks in nationally-representative survey data and related estimates from DiPrete, Gelman, McCormick, Teitler, & Zheng (2011). Results are intended to build upon both literature assessing same-race connectedness within associative networks, as well as literature exploring trends in social media use.

**BACKGROUND**

**Understanding Segregation Offline: The Roles of Structure and Agency**

Despite efforts to increase cross-race contact and promote diversity, racial segregation continues to be a defining characteristic of social networks (Hellerstein, Neumark, & McInerney, 2008; Moody, 2001; Quillian, 2002; Quillian & Campbell, 2003). In high school and middle school, adolescents are twice as likely to have a same-race friend as a cross-race friend (Moody, 2001). The tendency to associate with similar others is taken by some as a near social fact (DiPrete et al., 2011), and this tendency often manifests along racial lines. Though the frequency of contact between individuals of difference races is increasing (Sigelman & Welch, 1993), cross-race friendships are still relatively uncommon (Dunsmuir, 2013; Mouw & Entwisle, 2006; Sigelman & Welch, 1993).

Explanations of factors driving racial segregation within friendship networks generally highlight one of two factors: structural constraints – such as redlining or racial steering (Massey and Denton, 1993) or individual agency - such as a preference to make friends similar to oneself (McPherson et al., 2001). These two factors are not mutually exclusive. A number of studies showcase the relationship between structural- and agency-based factors in creating racially segregated associative networks by highlighting, for example, how residential segregation restricts individual friendship choices in schools by limiting cross-race exposure (Mouw & Entwisle, 2006). The following sections will provide an overview of research that examines structural and individual factors associated with persistent racial segregation across offline contexts in the US.

*Explaining Racial Segregation: The Structural Perspective*

The publication of *American Apartheid* in the 20th century (Massey & Denton, 1993) sparked a fresh wave of investigation regarding why segregation was so prevalent within the US, why this segregation persisted long after the formal dissolution of segregated spaces, and what sort of impact neighborhood-level segregation has had on the life outcomes for those within each segregated sphere. Examining how racial segregation persists on a macro-level invited scholars to consider factors that lead to the emergence and maintenance of racial segregation within associative networks as well. As a result, social scientists have dedicated a significant amount of effort toward understanding the structural mechanisms that lead to the development of segregated friendship networks.

The rise in popularity of suburban living and white flight from city centers influenced patterns of city-wide segregation, but this process was also influenced by institutional mechanisms such as redlining, racial steering and restrictive zoning (Massey & Denton, 1988, 1993; Muller, 1981). Despite the fact that overtly discriminatory housing and zoning policies have largely been dismantled, neighborhood segregation persists and the legacy of these polices still impact network formation. Many cities and towns remain heavily segregated, which reduces opportunities for same race contact through schools and everyday interactions. Massey & Denton (1988) highlight that as a result of these forces minority members of a population may be spatially distributed in a way that renders them underrepresented in some geographic areas and overrepresented in others. These patterns of overrepresentation and underrepresentation have unique behavioral and social implications - particularly in regards to individuals' interracial exposure and the resultant likelihood of individuals of different racial backgrounds being connected through the same social network. Furthermore, existing literature addresses the importance of neighborhood in helping individuals establish social ties and build social capital (Sampson, Morenoff, & Gannon-Rowley, 2002), meaning that segregated neighborhoods are in part responsible for building segregated networks.

Shifting the structural explanation of racial segregation in associative networks from the residential level, Moody (2001) seeks to explain what school structures promote segregation among students. He hypothesizes that the way in which classes and extracurricular activities are organized within the school directly impacts interracial contact, and in doing so likely accounts for within-network segregation. Incorporating both school structure and neighborhood structure into their explanation of racial segregation, Mouw & Entwisle (2006) examine why racial segregation persists in schools even after repeated attempts to diversify the student body. These authors find that approximately one third of the segregation found in schools is attributable to residential segregation. Students are more likely to be friends with other students within their immediate residential proximity (a distance of 0.25 km or less – a trend they refer to as the "bus stop effect"), and that and that those within their immediate proximity are usually of the same race. Despite the limited setting in which these analyses took place, all illustrate how structural forces guide patterns of interaction and consequentially influence the racial composition of individuals' social networks.

*Explaining Racial Segregation: The Agency Perspective*

In addition to understanding how structure influences the racial composition of individuals' networks, we can examine this phenomenon from an agency or individual preference level as well. McPherson et al. (2001) helped shift conversations about racial segregation within associative networks toward an agency-based perspective by emphasizing the tendency of individuals with similar characteristics to group together in a social setting. These authors state that homophily in social networks is exaggerated by factors such as family, propinquity, institutions, and isomorphic positions in social institutions, but it is nonetheless grounded in individual choices (McPherson et al., 2001). They find that rates of within-network racial diversity are generally much lower than would be expected if individuals chose connections at random. These findings suggest that in general, there exists low baseline homophily – homophily attributable to actual diversity in interpersonal exposure - and high inbreeding homophily – homophily attributable to individuals' preference to connect with similar others – within individual friendship networks.

Building on McPherson et al. (2001)'s proposal that individuals actively make the choice to befriend others who are like themselves, DiPrete et al. (2011) uses data from the 2006 General Social Survey (GSS) to determine if evidence of self-selected segregation appears not only along racial lines, but along dimensions such as political ideology, religious affiliation, and socioeconomic status (SES). These authors measure (1) how socially connected Americans are to one another (DiPrete et al., 2011: 235) and (2) how often these connections cross racial boundaries (among other factors). They found that Americans are highly segregated both within their core groups and among their acquaintances across all dimensions analyzed – including race - lending favor to McPherson et al. (2001)'s homophily proposal. In regard to race specifically, evidence from this study suggests that while segregation along racial lines did not increase between 1970 and 2000, it also did not decrease despite repeated efforts to create residential and educational conditions that enable cross-race friendship. These findings lend support to the idea that individuals often act upon the individual desire to seek similar friends.

Identity may also play a role in determining the racial composition of individuals' social networks. According to social identity theory, personal identity is heavily influenced by known membership within various social categories (Tajfel, 1982). Memberships within some categories, such as racial groups, are more salient to self-concept than others; for these influential categories, individuals tend to favor "in-group" interactions over those with "out-group" actors (Goar, 2007; Tajfel, 1982). Even as legal and political divisions between racial groups have diminished over time, racial identity remains a strong factor influencing whether and how often, members of different racial groups interact (Goar, 2007). As a result, racial identity continues to shape social interaction even in the absence of structural barriers separating members of different racial groups.

*Addressing the Interdependence of Structure and Agency*

Structure and agency influence network segregation in unique ways, but these factors are highly interdependent. While structure may influence cross-race contact, individual agency may influence who within that contact pool an individual may choose to befriend. A number of studies that examine racial segregation highlight this interaction. Mouw and Entwisle (2006)'s study of racial segregation within schools, for example, cites residential segregation as a structural factor that influences patterns of cross-race exposure among students, but nonetheless implies the role of agency in prompting students to *choose* friends based on residential proximity. Similarity, DiMaggio & Garip (2012) emphasize that residential segregation perpetuates facets of inequality directly related to race, such as SES. Residing and interacting within segregated settings impacts individuals' interpersonal exposure and influences their friendship choice by constraining their contact pool (DiMaggio & Garip, 2012). This sequence highlights the interdependent roles of structure and agency, in which individual agency is constrained by propinquity and the related social norms and expectations derived from this propinquity.

Regardless of the way in which prior research addresses the unique effects of both structure and agency, it is important to note that these forces are not separate and do not exist independently of one another. As stated by Martin & Dennis (2013) "there are no such 'things' as social 'structures,' 'classes', or indeed 'societies', yet terms such as these are indispensable, not only for sociologists but for the purposes of everyday communication" (14). This vocabulary provides a basis for understanding structure and agency as semi-stable entities. However, the fluidity and interdependence of structure and agency require researchers to understand how these forces influence opportunities for choice and behavior, especially in emergent contexts such as online social media spaces.

*Structure and Agency Online*

While the topic of racial segregation within social networks is well explored within offline spaces, little research has addressed whether comparable patterns of segregation exist within online spaces. Indeed, the key factors driving racial segregation offline (e.g., the historic legacy of discriminatory housing laws or a tendency toward homophily) may be non-existent or operate differently in social media spaces. It is difficult to anticipate the extent to which online networks would exhibit racial homogeneity. Furthermore, determining whether patterns of segregation are influenced by structural factors that shape exposure to other users or agency-based factors such as a preference for same-race connections requires significant unpacking.

There are some ways in which we might expect structural- and agency-based factors to foster patterns of segregation online that significantly differ from patterns that exist offline. Some social media – such as Facebook – may augment or parallel users' offline connections (Wimmer & Lewis, 2010). Twitter networks, on the other hand, may be composed of a mixture of users known offline and users known only through the website/platform (Duggan & Smith, 2016). This may minimize the effect of offline structural constraints such as residential segregation or administrative practices in schools. Additionally, it is difficult to predict how agency-based factors may operate within this space. Online spaces may be considered a "habitus of the new" (Papacharissi & Easton, 2013) where social structure in the forms of norms and expectations along with individual agency are continually being negotiated and redefined (Little, 2011; Martin & Dennis, 2013). This process of rapid co-evolution renders it difficult to predict whether agency-based factors influence tie selection in similar ways online as they do offline. Moreover, individuals may strategically hide and reveal components of their offline selves online, which may complicate users' ability to seek out or be found by users of the same race.

Despite the unique structure of online spaces and the apparent disconnect between online and offline social networks, it is also possible that the same structural- and agency-based factors that influence segregation offline have infiltrated online spaces. For instance, online users may choose to make connections with same-race friends as a way of confirming race as a salient social identity. In addition to this, existing literature suggests that online spaces are not immune to structural factors that influence racial inequality offline. Indeed, existing studies have shown that preferences for particular social media spaces are sometimes divided along racial lines (boyd, 2011). Finally, regardless of a perceived online/offline disconnect, offline geography may still influence online relationships which may create online/offline similarities in patterns of segregation.

**Racial Segregation Online**

*What Do We Know About Race and Friendship Online?*

This study examines the extent to which segregation may be observed within online networks. Furthermore, it seeks to understand whether patterns observed are generated by opportunity or choice. Analysis focuses on Twitter, a social networking site unique in its directed network relationships and sparse profile content. This paper examines the racial composition of egocentric user networks to compare the observed segregation of black and white users, unpack whether these patterns are driven by opportunity or by choice, and compare these patterns to known patterns of same-race connectedness in acquaintances offline.

A small body of existing literature has sought to address the role of race in online network development. Among the first of such studies were analyses of how race impacted users' preference for different social media platforms, broadly considering how users' demographic characteristics – including race – were associated with their choice of social networking sites. boyd (2011) used qualitative interviews to study how race played a role in users' decision to transfer their social media presence from MySpace to Facebook. Researchers have also considered how racial associations are displayed online. For example, Thelwall (2009) examined patterns of homophily within MySpace by analyzing the relationship between users' characteristics and their frequency of interaction with similar users within their network. This work suggests that structural and agency-based forces that influence segregation may continue to affect cross-race exposure and friendship selection within these spaces.

While existing literature attempts to draw upon the advantages of social media to examine racial segregation trends online, there are important dimensions of online interaction that these studies do not address. Studies such as Hargittai (2007) and boyd (2011) offer insight into how the racial composition of social media may influence friendship choices by limiting cross-racial exposure on a given site, but they do not examine observed patterns of same-race connectedness within these sites. In addition, most studies do not consider sites in which ties between users are symmetric/mutual or asymmetric/directed. Establishing a mutual tie is likely to be perceived as costlier than establishing an directed tie, and it is possible that this requires a more intimate connection or stronger sense of certainty to initiate. Finally, these studies primarily analyze sites that contain rich and informative user profiles. It is unclear whether such patterns will emerge within more anonymous spaces.

Given this current gap in our understanding of segregation within online spaces, this study seeks to examine same-race connectedness in Twitter, a social media space in which users' online networks may not necessarily parallel their offline networks, in which ties are directed, and in which profiles are sparse and variable. While it is likely that some level of segregation exists within this context, scholars have yet to examine what levels of segregation exist for users of different races on Twitter or predict whether these processes are driven by opportunity or choice. This study will address these questions and will also consider the challenges of: (1) sampling users from Twitter, (2) assessing users' demographic characteristics, (3) leveraging the diversity of tie relationships found on Twitter, and (4) determining which processes generated the patterns of same-race connections observed.

*The Twitter Environment*

Twitter is a microblogging platform that allows users to post single photos, videos or links and text-based content containing 140 characters or less. Messages – called "tweets" – tend to focus on personal updates, humor, or thoughts on media and politics.

This concise format allows users to update their blogs multiple times per day, providing a real-time depiction of their thoughts and experiences (Java, Song, Finin, & Tseng, 2007). In addition to projecting thoughts, users can communicate with one another through private messages, by re-tweeting a post from another user, or by using the @ command to reach out reply to a post from another user. They may also contribute to broader conversations by including a *hashtag* identifier in their tweet. Users are shown tweets from accounts that they follow in a feed that is updated in real time. Twitter was originally intended to be used via mobile devices, but tweets can also be sent and viewed using other internet capable devices, including tablets and personal computers.

Network structure on Twitter is different than that of other well-known social networking sites such as Facebook. Whereas Facebook is characterized by mutually acknowledged friend connections and often parallel or supplement users' existing friendship networks (Duggan & Smith, 2016; Vitak, Steinfield, & Ellison, 2011), the "Twittersphere" features directed network relationships that allows users to "follow" another user without considering whether that user will acknowledge or reciprocate the tie. The level of reciprocity in users' networks varies significantly according to how he or she intends to use the platform. Krishnamurthy, Gill, & Arlitt (2008) distinguish three types of Twitter users based on the structure of their associative networks.  One group – the broadcasters – is characterized by a large number of followers but a small number of reciprocated ties. Acquaintances are users who tend to establish mutual ties with their followers. Finally, there are those who have small follower networks but follow a large number of users either to spam them (miscreants) or to gain followers (evangelists). Although this taxonomy is a simplification of the types of unique personal network structures that exist on Twitter, it nonetheless represents what makes network formation on Twitter particularly diverse and unique.

*Anticipating Patterns of Segregation within Twitter*

Addressing the structure of networks and social ties on Twitter is critical to understanding why Twitter constitutes a unique space in which to study racial segregation. This structure suggests that Twitter users may not be subject to structural constraints that exist offline. According to previous research, macro-level, structural conditions – such as redlining or school tracking - play an essential role in understanding why patterns of racial segregation persist in other contexts (Massey & Denton, 1988b, 1993; Mouw & Entwisle, 2006). Thelwall (2009) summarizes: "in the offline world there are many [structural] factors that promote baseline homophily of various types so that people encounter others that are more similar to themselves in some way than average for the general population" (221). Given that an estimated 85% of Twitter users' networks are composed of users that they do not know personally or a combination of personal and non-personal connections (Duggan and Smith 2016), it is possible that structural factors perpetuating segregation are less influential or nonexistent within this space. Overall, while the offline world is structured along barriers and borders, Twitter is (in theory) relatively "borderless."

As discussed previously, notions of structure and agency, while often highlighted separately in literature surrounding patterns of racial segregation offline, are actually fluid and inseparable. This is especially true of social media, where users actively create and maintain a normative "structure," which reciprocally places constraints on their expressions of agency. We can frame an explanation of this condition within Twitter specifically in terms of *place* and *space*. As described by Harrison & Dourish (1996), *space* refers to the objective structure of an environment and *place* is what happens to spaces when users transform them into social settings with unique behavioral appropriateness, cultural expectations, and other normative constraints. Given this distinction, it is clear that Twitter not only differs from offline contexts in terms of *space* but also in terms of *place*. Within Twitter, the active, user-generated construction of *place* – an entity somewhat synonymous to Bourdieu's notion of the "habitus" – blurs the line between structure and agency and renders the influence of each on network segregation ambiguous (Taylor-Smith 2012). Thus, aside from the structural differences in space between the offline social world and the Twitter community, the co-evolution of structure and agency within Twitter invites social researchers to consider the phenomenon of racial segregation and investigate whether it persists given the norms and expectations within this unique and evolving social place.

Despite the blurred relationship between structure and agency online, it is possible that what social scientists know about agency and patterns of friendship segregation may not change as these connections move online. Twitter users may still prefer to follow and/or communicate with others who are similar to them, and factors promoting homophily may persist even when opportunities to interact with diverse peoples expand in online social media spaces. According to Sigelman & Welch (1993), the number of cross-race contacts in individuals' friendship networks is determined at least in part by preferences and biases developed in early childhood. It is possible, then, that these preferences and biases will carry into online spaces and generate online networks that are just as segregated as users' offline connections. In addition to this, it is possible that racial identity is such a salient social identity for users that they continue to seek out connections with others who affirm this identity within online spaces.

In addition to the influence of identity, existing research suggests that online spaces may not be immune to the social forces that guide patterns of segregation offline. boyd (2011), for instance, illustrates that offline forces may influence site choice and thus limit cross-race exposure within sites. Furthermore, while Twitter networks contain a more diverse mixture of known and unknown ties than Facebook (Duggan and Smith, 2016) some users may have networks that overlap heavily with their offline networks and thus carry racial homophily into this space. Overall, geography and distance may still matter in the age of the internet (Pflieger, Rozenblat, Mok, Wellman, & Carrasco, 2010), and structural forces that guide segregation may exist within both.

## METHODS

### Data

Data used for this study were collected using Twitter's REST application programming interface (API) - a quick but limited access point for gathering behavioral trace data directly from Twitter. The REST API allows researchers to access tweets created within the past nine days, core information about user accounts – including handles, profile photos, location, and more – as well as users' connections, interactions, and the timelines (or the aggregated text content of their profiles). Our data includes a sample of US based Twitter users and includes metadata for each user and the users they follow/are following. Throughout this analysis, we will refer to this set of sampled U.S. based Twitter users as *egos*, and those following and/or followed by the egos as *alters*. *Figure 1* provides an illustration of the data.

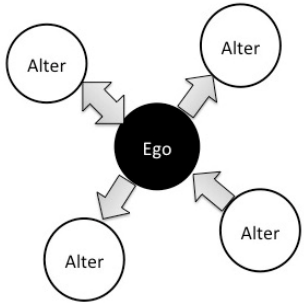

*Figure 1:* Network Structure

Preliminary exploration of Twitter user IDs indicate that they are assigned in a largely linear and monotonic manner, but it is unclear whether each ID value is assigned sequentially. Given this, we use a keyword-based sampling approach that attempts to randomly select active users based on usage of randomly selected common keywords from Ogden's Basic English word list (see: http://ogden.basic-english.org/). This method yielded a sample of 347,990 ID values from which the egos in our dataset were sampled. In total, we sampled 25,000 IDs at random from this set and successfully identified 23,150 of these ID values as those for active accounts.

In order to situate our findings within existing literature examining race within associative networks, we reduce our sample to include only users from the US. In order to do this, we rely upon the location data listed within users' Twitter profiles. Although location is an optional profile field, many users choose to provide some indication of where they live. Because the content of the field varies greatly, this data does not lend itself well to automated analysis. For example, a user living in Washington state may list "Pacific Northwest" or "PNW" as their location, and a user from California may list "The Golden State." Given this, we used Amazon's Mechanical Turk (AMT)[1] workers to identify whether a user's listed location is a location in the US. This yielded 15,594 accounts with information in the location field, 4,245 of which were confirmed by AMT workers to live within the US.

---

[1] AMT is a marketplace for work that requires human intelligence and offers an efficient way to obtain answer to specific questions by defining new tasks (known as a Human Intelligence Tasks or HITs), which can be performed by online workers. AMT workers, or AMT workers, are anonymous, independent individuals who are identifiable only by their unique ID numbers. Previous research indicates that the AMT workers are skilled at a variety of tasks and are highly reliable experimental research subjects (Buhrmester, Kwang, & Gosling, 2011; Mason & Suri, 2012; McCormick, Lee, Cesare, Shojaie, & Spiro, 2015) as well as skilled and demographically diverse survey respondents (Behrend, Sharek, Meade, & Wiebe, 2011).

In order to draw comparisons between these results and existing literature regarding racial segregation within associative networks, only black and white users were selected for analysis. Using methods outlined by McCormick et al. (2015), we again used AMT workers to evaluate the race of the egos. Under this framework, each photo was displayed to three AMT workers, who were asked to estimate the race (i.e. black, white, Asian, other or unknown) and ethnicity (i.e. Hispanic/Latino or non-Hispanic/Latino) of the primary individual in the photo. AMT workers were also asked to identify whether the photo displayed an image of a person, a group of people, or something other than a person (a logo, pet, object, etc.). While we acknowledge that these estimates are based on others' impressions and in some cases be inconsistent with personal identity, research has confirmed that the AMT workers nonetheless provide estimates of users' approximate age, sex and race that correspond well with estimates provided by expert trained coders. This left us with a total of 561 black egos and 1,841 white egos to include in this analysis. We used Twitter's REST API to collect and save the ID values and metadata of each ego's friends and followers.

**Extracting Demographic Information from Twitter Profiles**

We use profile photo data to infer the demographic characteristics of users – egos and alters. Due to the cost and time required to infer demographic characteristics using AMT, we choose to probabilistically subsample the ties within each ego's network. Given that preliminary analyses indicated that the minimum non-zero same-race connectedness of egos is approximately 12%, we determined that we were 95% likely to find at least five different-race friends by sampling approximately 65 ties (alters) per ego. Given this, for all egos with a count of unique friends and followers between 65 ties and the 95th percentile of tie counts (17457 ties) we randomly selected from their networks 65 unique alters. We also removed all egos with an alter count above the 95th percentile, and retained the full networks of egos with 65 alters or fewer.

After collecting and sub-sampling network data, we sample within our egos. In order to a.) ensure we are likely to detect a true average same-race connectedness difference between these groups of five percent or more and b.) render the inference of demographic information from these photos as efficient and cost effective as possible, we used a random quota sample to select 460 white and 460 black egos to include in the analysis. The network characteristics of the users included in this analysis (with full network ties rather than randomly subsampled ties) are displayed in *Table 1*.

*Table 1:* Characteristics of Ego Networks (not randomly subsampled)

| Ego race | | Follower | Following | Mutual |
|---|---|---|---|---|
| Black Users | Mean (SD) | 1436 (1918) | 1016 (1183) | 619 (967) |
| | Min | 7 | 0 | 0 |
| | Max | 16759 | 10061 | 9535 |
| White Users | Mean (SD) | 1375 (2002) | 1054 (1457) | 615 (1196) |
| | Min | 0 | 0 | 0 |
| | Max | 13217 | 11316 | 9502 |

The race of the subsampled alters for these egos was estimated using a two-step process. First, drawing on the precedent of Zagheni, Garimela and Weber (2014), the profile image URLs of the alters were run evaluated using the facial recognition software API Face++[2]. Only results for photos that were estimated with 99 percent confidence or higher were retained. For photos featuring multiple faces, results were retained only if all users were estimated to be the same race, and the average confidence of the estimations met or exceeded 99 percent. This method captured approximately 9% of the total alters used. For the remaining alters, race was coded using evaluations from AMT workers and methods outlined by McCormick et al (2015). In total 98.6% of all alter photos were successfully coded.

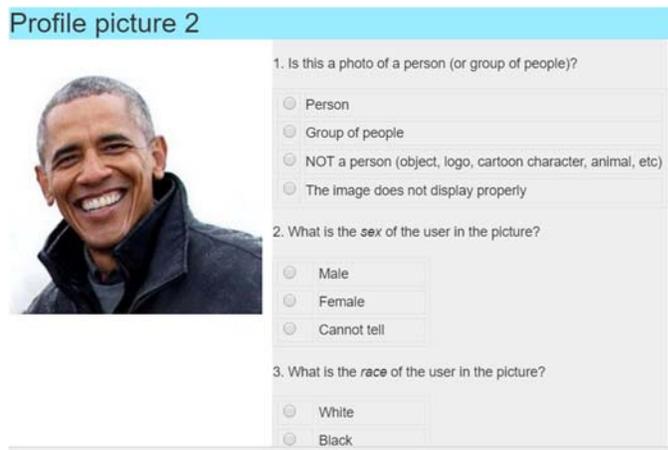

*Figure 2:* Example of coding task from Mechanical Turk

## RESULTS

This analysis is divided into three parts. The first part provides a descriptive illustration and comparison of same-race connectedness among black and white egos. The second part seeks to unpack whether the patterns of segregation observed are generated by opportunity or by choice. The third part contextualizes findings by comparing measures of segregation within sampled Twitter networks to data from a 2006 GSS module that asks respondents to estimate the proportion of their acquaintanceship ties that are the same race as themselves.

Within the context of this analysis, alters with the racial categorization "unknown" are treated as a separate racial category. Some alters may represent a brand or entity, and some may be private individuals who wish to present a version of themselves within Twitter that does not accurately reflect their racial identity (Nakamura, 2013). Because some alters whose photos are categorized as "group" or "person" lack clear racial categorization due to picture quality or content and thus fall into the 'unknown,' category, we perform sensitivity analyses for portions of these findings that classify these users as either all white or all black (see *Appendix B*).

---

[2] Face++ offers a public API for Automated Facial Recognition. See www.faceplusplus.com for additional details.

As mentioned previously, Twitter is unique from sites such as Facebook in that it features directed ties. On Twitter, a user may elect to follow another user without consideration of reciprocation. This unique structural trait facilitates a variety of usage patterns and friendship types on Twitter. As described by Krishnamurthy et al. (2008) friendship patterns on Twitter appear to be motivated by either a desire to establish interpersonal connections (termed by Krishnamurthy et al. As 'acquaintances') or a desire to gather and/or spread information (termed 'evangelists'/'miscreants' and 'broadcasters', respectively). This study seeks to leverage this unique structural feature of Twitter and analyze patterns of segregation by tie type. The following analyses consider the ego's outgoing ties – those whom the *ego is following* - as an expression of friend preference. It treats *mutual* ties between ego and alters as the strongest available connection – something akin to acquaintanceship within this space.  It acknowledges that while incoming ties – those who *follow the ego* - reflect very little about choice on the part of the ego, they nonetheless provide important context about the ego's broader network.

**Black and White Same Race connectedness by Tie Type**

To help understand whether the openness of Twitter helps facilitate diversity within users' networks, we first compare the same-race connectedness of black and white egos by tie type (i.e. all unique alter ties, alters followed by the ego, alters following the ego, and mutual connections between the ego and alter). Examining these patterns by tie type lends valuable granularity to our understanding of same-race connectedness within Twitter, as tie directionality may be related to the strength and nature of the relationship between ego and alter. Results of this analysis are shown in Table 2. Observed distributions are displayed in blue in *Figure 3*.

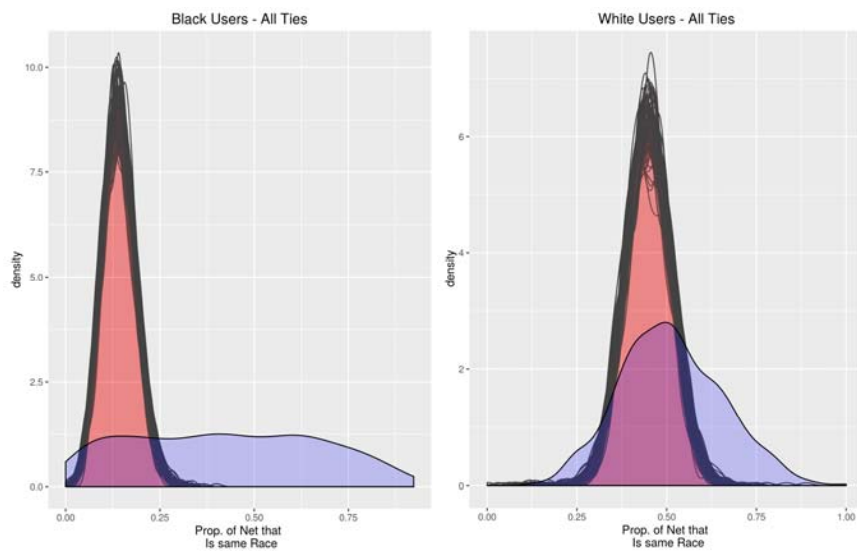

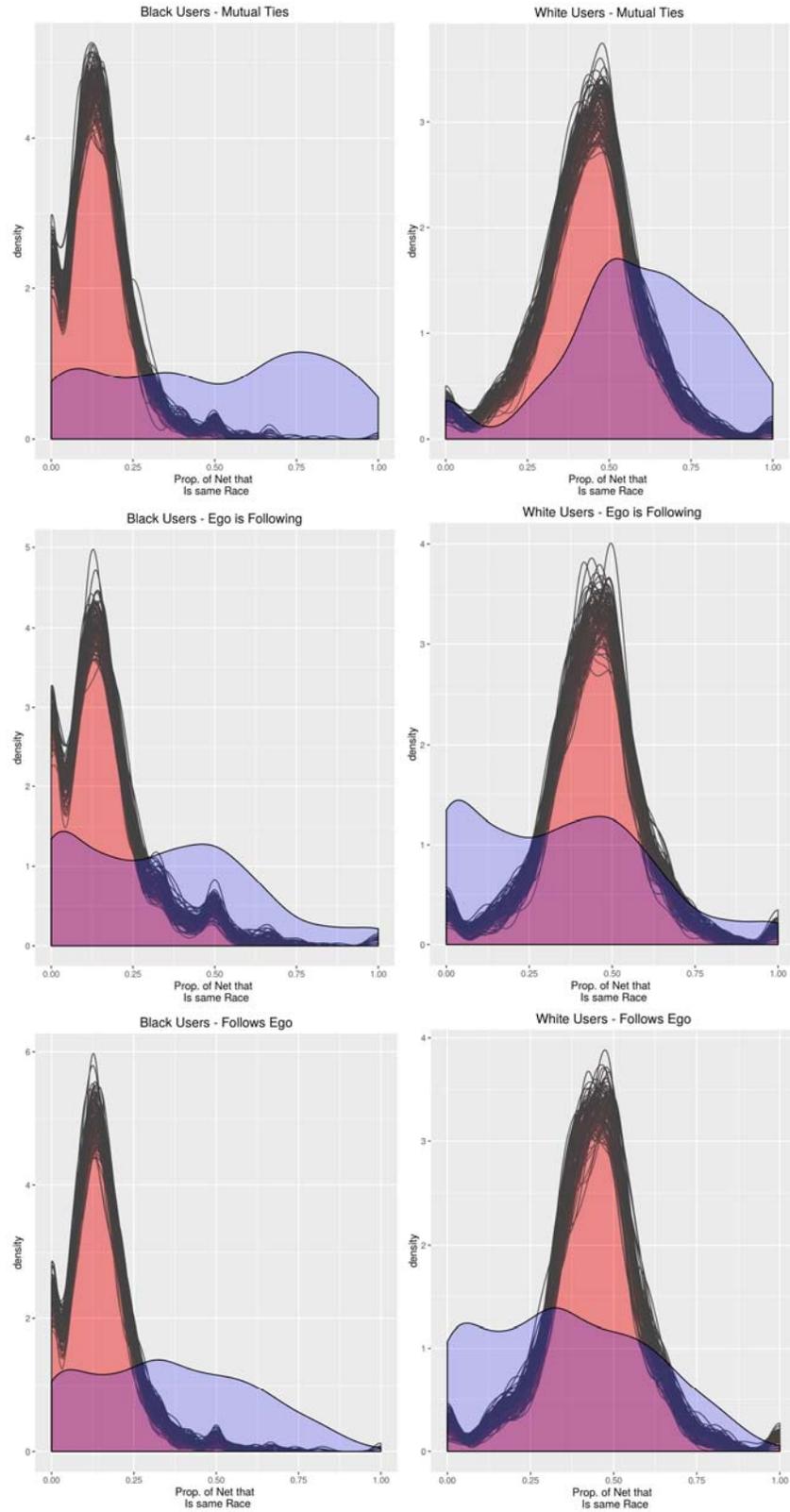

*Figure 3*: Black and White Egos Same-Race Connectedness by Tie Type
(Blue=Observed, Red=Simulated)

*Table 2:* Same-Race connectedness for black and white egos by tie type

|  | Mean | SD | Lower quart | Upper quart | Min | Max |
|---|---|---|---|---|---|---|
| **Black egos** |  |  |  |  |  |  |
| Total ties | 0.423 | 0.246 | 0.206 | 0.626 | 0.000 | 0.923 |
| Mutual | 0.498 | 0.316 | 0.200 | 0.778 | 0.000 | 1.000 |
| Following | 0.340 | 0.268 | 0.106 | 0.500 | 0.000 | 1.000 |
| Follower | 0.358 | 0.247 | 0.145 | 0.550 | 0.000 | 1.000 |
| **White Egos** |  |  |  |  |  |  |
| Total ties | 0.512 | 0.141 | 0.415 | 0.616 | 0.143 | 0.892 |
| Mutual | 0.597 | 0.239 | 0.467 | 0.769 | 0.000 | 1.000 |
| Following | 0.446 | 0.218 | 0.307 | 0.583 | 0.000 | 1.000 |
| Follower | 0.461 | 0.201 | 0.333 | 0.600 | 0.000 | 1.000 |

Viewing the distribution of black and white same-race connectedness by tie type observed on Twitter (see blue distributions in *Figure 3*) we note that the same-race connectedness of white egos generally follows a normal to slightly positively skewed distribution, but the distribution of same-race connectedness among black users is more uniform across tie types. These striking visual differences indicate that the way in which users experience Twitter as a diverse social space varies by race. T-test comparisons of the average same-race connectedness of black and white egos (displayed in *Table 2*) indicate that there are strong black/white differences in same-race connectedness across all tie types. The difference in same-race connectedness between black and white egos appears be strongest when considering those whom the ego follows. Under this criteria, black egos follow significantly fewer same-race connections than do white egos (t=-7.44, p<0.001). Black egos also have significantly fewer same-race mutual ties (t=-6.453, p<0.001), same-race followers (t=-5.71, p<0.001), and total same-race connections (t=-11.622, p<0.001). Sensitivity analyses in Appendix A reflect these trends as well.

Given the unexpectedly uniform distribution of observed same-race connectedness for black egos, we elect to engage in a close, qualitative analysis of egos on either extreme of the homophily distribution for total ties. Examining the ten black egos with the most and least homophilous networks, we note significant disparities in Twitter use. Egos with remarkably low levels of homophily appear to primarily represent older users and users who engage with the Twitter audience as a means of promoting a personal business or brand. Those with high levels of homophily, however, appear to use Twitter as a purely social platform, providing personal details and updates on their personalities and lives.

**Opportunity and Choice within Friendship Networks**

While finding black/white differences in same-race connectedness for egos in this sample indicates that there is a racial divide in how users experience the potential diversifying effects of Twitter, these patterns could be driven by the fact that black users are still a minority on Twitter and are less likely than white users. Given this, compared the distribution of same-race connectedness observed among egos to the distribution of same-race connectedness we would expect if they chose friends without regard to race. This lends depth to the analysis by lending insight into whether users select same-race friends due to propinquity or choice.

This analysis acknowledges the challenge of defining opportunity in Twitter due to the fact that a.) Twitter algorithmically suggests users to follow and b.) it is unclear to what extent users' offline and Twitter social connections overlap. However, it conceptualizes Twitter as a bounded social space and examines interactions only as they occur within this space. Furthermore, it assumes that in absence of a preference for homophily, users will select connections from a pool of other users with a given racial distribution.

To detect whether egos choose connections through opportunity or choice, we begin by simulating what users' networks would look like if they selected friends at random. Within each tie type for each user, we generate 100 simulated networks that select friends according to their estimated proportion within the Twittersphere. For instance, if the Twittersphere is approximately 45% white, then the expected probability of selecting a white friend is 0.45. We then compare the distribution of these simulated networks to the distribution of observed same-race connectedness within our data. Results of this analysis are displayed in *Figure 3;* red distributions represent simulated ties and blue distributions represent observed ties. If observed data deviates from these simulated distributions, then we may assume that observed same-race connectedness is the result of choice rather than opportunity.

*Table 3*: Mean Same Race connectedness among observed, simulated networks

|  | Observed | Simulated |
|---|---|---|
| Black egos | | |
| Total ties | 0.423 | 0.142 |
| Mutual | 0.498 | 0.145 |
| Following | 0.340 | 0.160 |
| Follower | 0.358 | 0.144 |
| White Egos | | |
| Total ties | 0.512 | 0.448 |
| Mutual | 0.597 | 0.439 |
| Following | 0.446 | 0.438 |
| Follower | 0.461 | 0.441 |

When considering users' total connections, the observed and simulated same-race distribution of white egos are both normal with similar means, although the observed data displays a slightly wider standard deviation than the simulated data. The observed and simulated distribution of all ties for black users, however, is markedly different. While the simulated distribution is normal with a mean near the estimated total proportion of black users within the Twitter population, the observed distribution of same race connectedness is notably uniform. These results alone suggest that many black Twitter users are more likely than white users to seek out and establish connections with same-race others.

When observing mutual ties, it seems that both black and white egos have a tendency to prefer same-race connection. For black and white egos, the observed same-race connectedness of mutual ties displays a more positive skew than the simulated data. Again, this effect seems stronger for black egos, as the simulated distribution appears normal and the observed data more uniform. When considering asymmetric (i.e. following and follower) relationships, Twitter seems to have a diversifying effect for white egos. Often times these egos have fewer same-race friends than we would expect if they chose friends at random. This is not the case with black egos, who seem to have on average more same-race connections than would be expected if they chose friends without regard to race. Taken together, these results suggest that even though Twitter opens the opportunity to create low cost networks, establishing ties – especially 'close' mutual ties – with similar others remains important for many black egos.

To further understand how choice operates to influence network composition, we draw upon the capabilities of the Twitter API to examine the sequence in which users select ties. It is expected that the Twitter API returns friendship connections in the approximate order in which they are established. Given this, we are able to view whether there is an association between the sequence of users' connections and the race of these connections. Primarily, we are interested in viewing whether users initially give preference to same-race connections and then diversity.

To examine this, an ego's connections are coded as "1" if they are the same race as the ego and "0" if they are not. If these connections are plotted by sequence, then a user who initially exhibits a strong preference for same-race connections but whose network diversifies over time would display a negative correlation (see *Figure 4)*. We may think of initial homogeneity may be a reflection of the user's unfiltered preferences, and the diversification of their network over time may emerge as a consequence of finding connections through others.

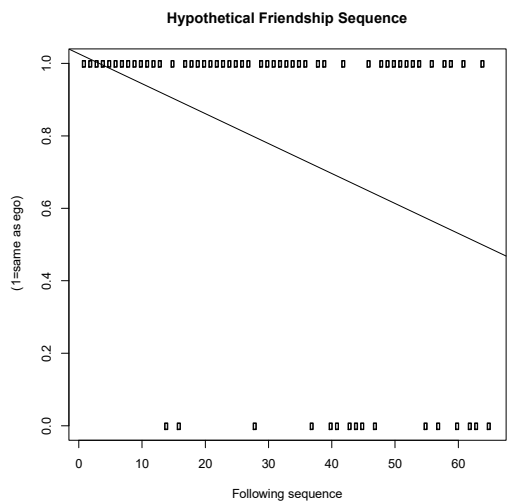

*Figure 4:* Friendship ordering sequence for hypothetical user.

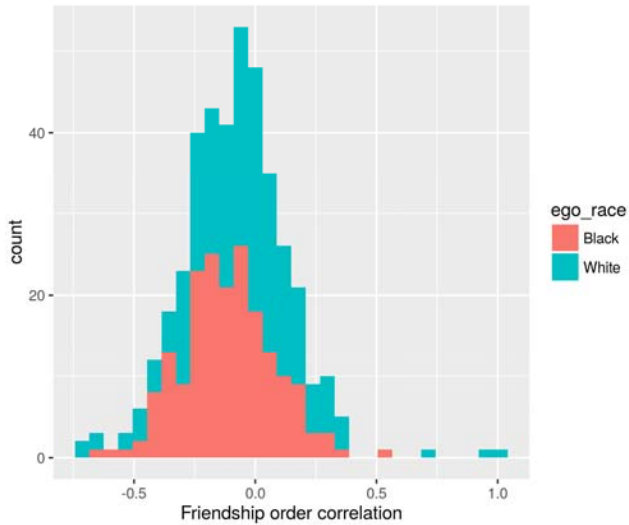

*Figure 5:* Distribution of friendship order/correlation by ego race

We find, however, that sequence does not play a strong role in friendship choice. Indeed, the average correlation between same-race connection and friending sequence is -0.11 for black egos and -0.05 for white egos. This indicates weak initial inclinations toward homophily for egos of both races.

**Comparing Measures of Segregation on Twitter to Measures of Segregation Offline**

The previous sections of this analysis illustrate racial differences in same-race connectedness within Twitter and provide some indication of whether Twitter users actively seek out same-race connections on this site. To help ground these findings within existing research regarding racial segregation in associative networks, we also consider whether the racial composition of Twitter networks parallels that of offline associative networks. We compare measures of same-race connectedness among sampled egos to measures of same-race connectedness among offline acquaintances as reported by the 2006 GSS special module on social connectedness.

The 2006 GSS module used in this analysis asks respondent to estimate the number of friends they have that are of their own race on a scale from 1 (almost all of them) to 5 (almost none of them). In contrast to the exact, observed measures in the Twitter data, the measures provided by the GSS acquaintanceship data are ordinal. In order to draw rough comparisons between the two, however, we consider the 1 to 5 measures to be equivalent to a 0-100% scale. *Table 4* illustrates the average percent of same-race acquaintances for respondents of each race calculated using midpoint estimation (i.e. 5=90%, 4=70%, 3=50%, 2=30%, 1=10%) (N=1206).

*Table 4*: How many of your acquaintances are of the same race as you?

| | Almost all are same (1) | Most are same (2) | About equal (3) | Most are not same (4) | Almost none are same (5) | Average Score Average. | Percent approx. using midpoint estimation |
|---|---|---|---|---|---|---|---|
| Black | 48 | 48 | 66 | 14 | 5 | 2.34 | 63.3 |
| White | 379 | 459 | 162 | 16 | 9 | 1.84 | 73.1 |

*Note: Table 4 displays the responses from the GSS to the question "Are the people that you are acquainted with 1.) Almost all the same race as you 2.) Mostly the same race as you 3.) About evenly divided between the same race as you and other races 4.) Mostly a different race than you 5.) Almost all a different race than you"*

In order to draw a comparison between the measures reported by the GSS and data gathered from Twitter, this section focuses exclusively on egos' mutual ties. While it is difficult to unpack the nuance and significance of friendships on Twitter, we may expect that mutual connections within this space are in some way similar to offline acquaintanceships - which the GSS defines as people whom the respondent would "stop and talk at for at least a moment" if they "ran into the person on the street or in a shopping mall."

In this analysis we manage alters of "unknown" race in two ways. The first analysis (*Table 5*) treats "unknown" as a separate category from which egos may choose when selecting friends. The second analysis (*Table 6*) proportionally assigns the "unknown" users a racial category according to the racial composition of Twitter as estimated by the racial distribution of randomly sampled egos.

Results in *Tables 5-6* indicate that if "unknown" cases are treated as a separate racial category from which egos can select acquaintances then both black and white networks are significantly less segregated than they are offline. The effect for black egos, however, is less strong. However, if we proportionally assign "unknown" cases a racial identity, this difference changes dramatically. White Twitter acquaintanceship networks appear to be *more* segregated than offline networks under this adjustment. Black networks remain less segregated, but this effect is less strong than what is observed in *Table 5*.

*Table 5*: Average percent of each individual's acquaintanceship network that is the same race as them ("unknown" as separate category)

| | GSS (percent equivalent) | Twitter |
|---|---|---|
| White | 73.1 (72.1 ,74.1) | 62.6 (60.4, 64.7) |
| Black | 63.2 (60.9, 65.6) | 47.3 (48.2, 53.0) |

*Table 6*: Average percent of each individual's acquaintanceship network that is the same race as them ("unknown" removed)

|  | GSS (percent equivalent) | Twitter |
|---|---|---|
| White | 73.1 (72.1 ,74.1) | 80.5 (78.7, 82.3) |
| Black | 63.2 (60.9, 65.6) | 54.9 (52.3, 57.6) |

The contents of *Table 5* suggest that Twitter does open the possibly of expanding the diversity of users' networks. However, the desegregating effect of Twitter appears to impact white egos more strongly than black users. However, when combined with the results of *Table 6* we note that this desegregating effect may have more to do with a preference among white egos to connect with "unknown" accounts, which may represent entities, brands, and news sources rather than individuals.

**DISCUSSION**

This project seeks to contribute to broader understanding of patterns of racial segregation in associative networks within Twitter – a social context that may be somewhat disconnected from offline structural forces and features evolving usage norms. While we expected to see some evidence of segregation within this space, the way in which structure and agency are co-evolving within Twitter made it difficult to anticipate the level of segregation or predict whether patterns observed are generated by opportunity or choice.

We begin by noting that black and white users have significantly different overall levels of same-race connectedness within Twitter. Paralleling what offline data documents – in which white individuals have very high levels of same-race connectedness but black individuals display more diverse networks (Dunsmuir, 2013) – the white egos analyzed have overall higher levels of same-race connectedness than black egos. This pattern holds true for all tie types on Twitter, and is consistent across sensitivity tests included in this analysis.

Central to this analysis is our desire to analyze whether patterns of segregation observed within the data are more likely driven by opportunity or choice. To answer this, we simulate what egos' networks would look like if they selected friends at random. We note that the same-race connectedness of black and white users is consistently more negatively skewed than that of the simulated distributions, indicating that users do elect to select same-race friends within this space. This effect seems to be strongest among black egos, particularly in regard to their selection of mutual ties. We do not find a strong association between the ordering of users whom an ego chooses to follow, and we find weak evidence that users have a preference to prioritize the selection of same-race ties (i.e., to connect with similar users first and diversify their network later). Overall, these results suggest that homophily does influence patterns of same-race connectedness within Twitter, despite the fact that Twitter offers the opportunity to diversity one's network at little social cost. However, this effect is strongest among black egos, and is most likely to manifest among mutual rather than directed ties.

Perhaps most interestingly we note that the tendency toward homophily varies significantly among black users. Among black egos analyzed the distribution of same-race connectedness across all ties and mutual ties is strongly uniform. This indicates that usage patterns vary significantly for black egos. While some users display highly diverse networks, others exhibit very high levels of same-race connectedness. A qualitative analysis of profiles among black egos indicates that many of the users who exhibit very low levels of same-race connectedness are either older users, or use their Twitter presence for self-promotion. Those with very high levels of segregation appear to use Twitter for purely social purposes. Overall, it appears that while homophily may exist on Twitter, its influence varies according to who is using the platform and for what purpose.

To help contextualize these findings within existing literature examining racial segregation in associative networks, this analysis also compares levels of segregation seen within Twitter to patterns of within-network segregation seen offline. Results indicate that despite the fact that individuals on Twitter still seem to exhibit a preference for same-race connections, they nonetheless have networks of mutual ties – or what we may think of as 'acquaintances' on this site that are more diverse than we see offline. This holds true for black users regardless of how "cannot tell" cases are treated in the analysis. For white users, however, it only holds true if "cannot tell" is treated as a racial category that is unique to Twitter. These results suggest that in comparison with offline contexts Twitter may encourage users to reach out to others who do not outwardly belong to the same racial category as them, but that this effect is stronger for white users than for Black users. Some studies suggest that Twitter is a context for information gathering more so than social interaction (Duggan & Smith, 2016; Kwak, Lee, Park, & Moon, 2010), so this difference may be attributable in part to a tendency among white users to engage with Twitter in this way.

Overall, these measures indicate that Twitter may indeed constitute a "habitus of the new" in which race places a less significant role in influencing users' friendship choices. However, this freedom to choose friends of a different race appears to impact the network composition of white users more so than black users. It is possible that this unequal impact is due to the fact that for black users, race is a particularly salient social identity that remains so even within this new context (Rowley et al. 2008). This may influence black users to be more likely to a.) display their race to other users and b.) seek out friends who are of the same race as them. It may also influence them to seek out and interact with others who share their concerns and experiences regarding race, making the observed difference potentially an artifact of "Black Twitter" (Clark, 2014). In addition to this, it is possible that black users' networks are more strongly connected to their offline networks than those of white users, and that this online-offline connection increases the influence of structural forces that perpetuate segregation within this space.

**CONCLUSION, LIMITATIONS AND DIRECTIONS FOR FUTURE WORK**

This paper suggests that while Twitter may be a "habitus of the new" where existing norms – including those regarding the role of race in the development of interpersonal connections – segregation and homophily nonetheless exists within this space. Specifically, we note that many black users have higher levels of same-race connectedness than we would expect if they chose friends without regard to race. While they on average experience lower levels of same-race connectedness than white users, and have more diverse networks on Twitter than they report offline, we nonetheless see evidence of homophily. For white users, Twitter may have a somewhat diversifying effect, but results comparing Twitter networks to measures from the 2006 GSS suggest that this may be driven in part by a preference to connect with brands and other non-person entities on Twitter.

Online social media spaces, such as Twitter, are revolutionizing the ways in which social scientists are able to examine the social world. Social media sites can be viewed as emerging social contexts that provide social scientists new avenues in which to examine the role of social structure and norms on behaviors and attitudes. Despite its potential, social media data present challenges to social science research for a number of reasons. For one, these data are dynamic; not only must researchers be aware of changes in the way the platform stores information, but must account for the fact that users' profiles and networks are constantly changing. In addition to this, different platforms have different affordances that restrict or permit the collection of particular types of user metadata. Additionally, social media spaces feature evolving patterns of use. Social scientists are responsible for tracking these changes and assessing the ways in which the current normative/cultural state of a social media space may impact their results.

The use of social media data to analyze the social world also raises unique ethical challenges. For one, because Twitter users as research subjects do not 'participate' in the research process in a traditional sense (Ang, Bobrowicz, Schiano, & Nardi, 2013), researchers must reevaluate and redesign institutional review board (IRB) procedural standards regarding the ethical treatment of research subjects. While procedures such as informed consent are not realistic when handling datasets of tens of thousands of users, this does not mean that those managing, producing and analyzing the data should not consider how to best to mitigate harm, ensure beneficence and protect privacy. Furthermore, ethical data use may relate to the manner in which results are framed and disseminated.  As boyd & Crawford (2012) suggest, the practice and dissemination of Big Data research raises important questions about truth, power and control. It is the responsibility of the researcher to consider the relationship between the results of a study and the perpetuation of inequality.

The analysis of segregation within Twitter raises some unique methodological challenges. For example, not all users choose to display their race through their profile picture and it unclear how researchers should characterize these photos. In this study we treated individuals who choose not to display their race through their profile photo as belonging to their own category, albeit a category in which racial identity does not play a role. While we believe this serves our analyses well, future research may incorporate other metadata to estimate users' race. Additionally, while the automated detection of social media users' gender is well-addressed within existing literature (Burger, Henderson, Kim, & Zarrella, 2011; Liu & Ruths, 2013; Mislove, Lehmann, Ahn, Onnela, & Rosenquist, 2011), estimating users' race or ethnicity with great accuracy remains a challenge. This study relies on human evaluation to assess the race of egos and alters. While this method ensure estimates are accurate and reliable (McCormick et al. 2015), it nonetheless requires us to scale down the size of the sample used.

Another methodological challenge associated with the user of Twitter data stems from the fact that Twitter is a dynamic data source and that the structure of the site and content of users' profiles can and do change over time. The way in which Twitter tracks users and stores user information is not static and it is the responsibility of the researcher to keep up-to-date with changes to Twitter and the Twitter API. In addition to this, users change the content of their profiles quickly, so it is important to examine photo-based metadata as soon as possible following data collection. Further, it is the responsibility of those who use these data to develop sustainable methods of preserving snapshots of user metadata – including profile photos – over time.

There are important limitations associated with these findings. For one, we recognize that these results cannot disentangle the influence of Twitter's friendship selection algorithms from the patterns of same-race connectedness observed. If these algorithms suggest friends that are similar to those already followed by or following the ego, then they may help perpetuate processes of segregation or diversification already in place. We do not believe this process strongly influences our results given that there seems to be little connection between friendship ordering and tie race, but it is nonetheless important to knowledge. We also cannot address the possible overlap between users' offline and Twitter connections. While Duggan & Smith (2016) estimate that only a small fraction of Twitter users (15%) have pre-existing relationships with their Twitter ties, the presence of any online-offline network overlap has the potential to perpetuate forces known to influence patterns of same-race connectedness observed offline within this space.

Through future research we hope to not only examine egos' mutual follower networks, but networks of users with whom the ego communicates as well. It is possible that networks of communication wherein users mutually engage in conversation through private messaging or public interaction – which may be thought of as conceptually equivalent to networks of friendship or trust – are not as diverse as users' tie-based networks. Narrowing the focus of our analysis to users who mutually communicate with one another may change the levels of segregation reported within this study.

**Appendix: Sensitivity Analyses**
# Sensitivity analysis set 1: Comparing same-race connectedness

*Sensitivity analysis 1A: All 'unknown' photos featuring people are assumed to be black*

Distribution of same-race connectedness for black and white egos

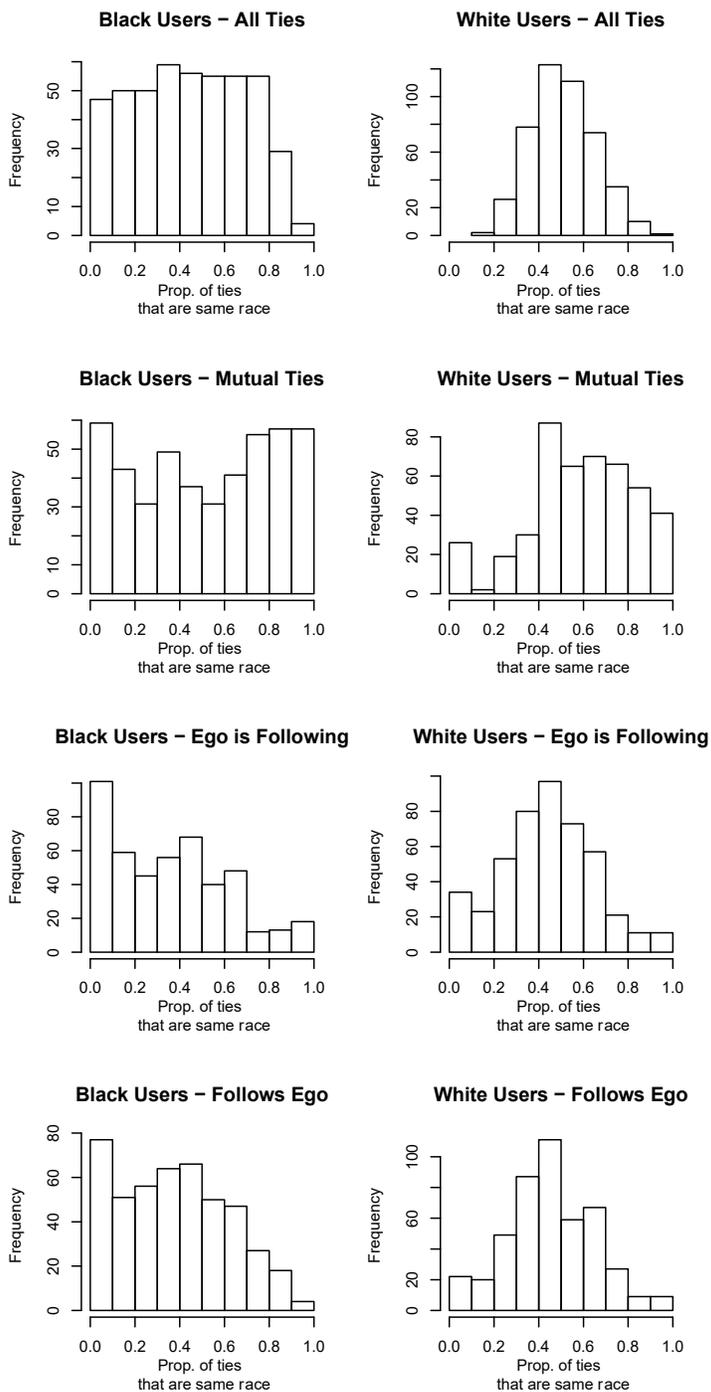

Same race connectedness: Black Egos

| Tie type | Mean | SD | Lower quart | Upper quart | Min | Max |
|---|---|---|---|---|---|---|
| Total ties | 0.448 | 0.245 | 0.234 | 0.651 | 0.000 | 0.923 |
| Mutual | 0.524 | 0.314 | 0.250 | 0.800 | 0.000 | 1.000 |
| Following | 0.383 | 0.272 | 0.182 | 0.571 | 0.000 | 1.000 |
| Follower | 0.363 | 0.248 | 0.195 | 0.5544 | 0.000 | 0.000 |

Same race connectedness: White Egos

| Tie type | Mean | SD | Lower Quart | Upper Quart | Min | Max |
|---|---|---|---|---|---|---|
| Total ties | 0.512 | 0.141 | 0.415 | 0.616 | 0.143 | 0.892 |
| Mutual | 0.597 | 0.239 | 0.467 | 0.769 | 0.000 | 1.000 |
| Following | 0.446 | 0.218 | 0.307 | 0.583 | 0.000 | 1.000 |
| Follower | 0.461 | 0.201 | 0.333 | 0.600 | 0.000 | 1.000 |

Mean B/W difference total: $t = -4.8718$, $df = 733.58$, p-value = 1.356e-06
Mean B/W difference mutual: $t = -3.9793$, $df = 857.47$, p-value = 7.495e-05
Mean B/W difference following: $t = -5.0829$, $df = 876.84$, p-value = 4.545e-07
Mean B/W difference follower: $t = -5.3073$, $df = 880.24$, p-value = 1.409e-07

*Sensitivity analysis 1B: All person unknowns are white*

Distribution of same-race connectedness for black and white egos

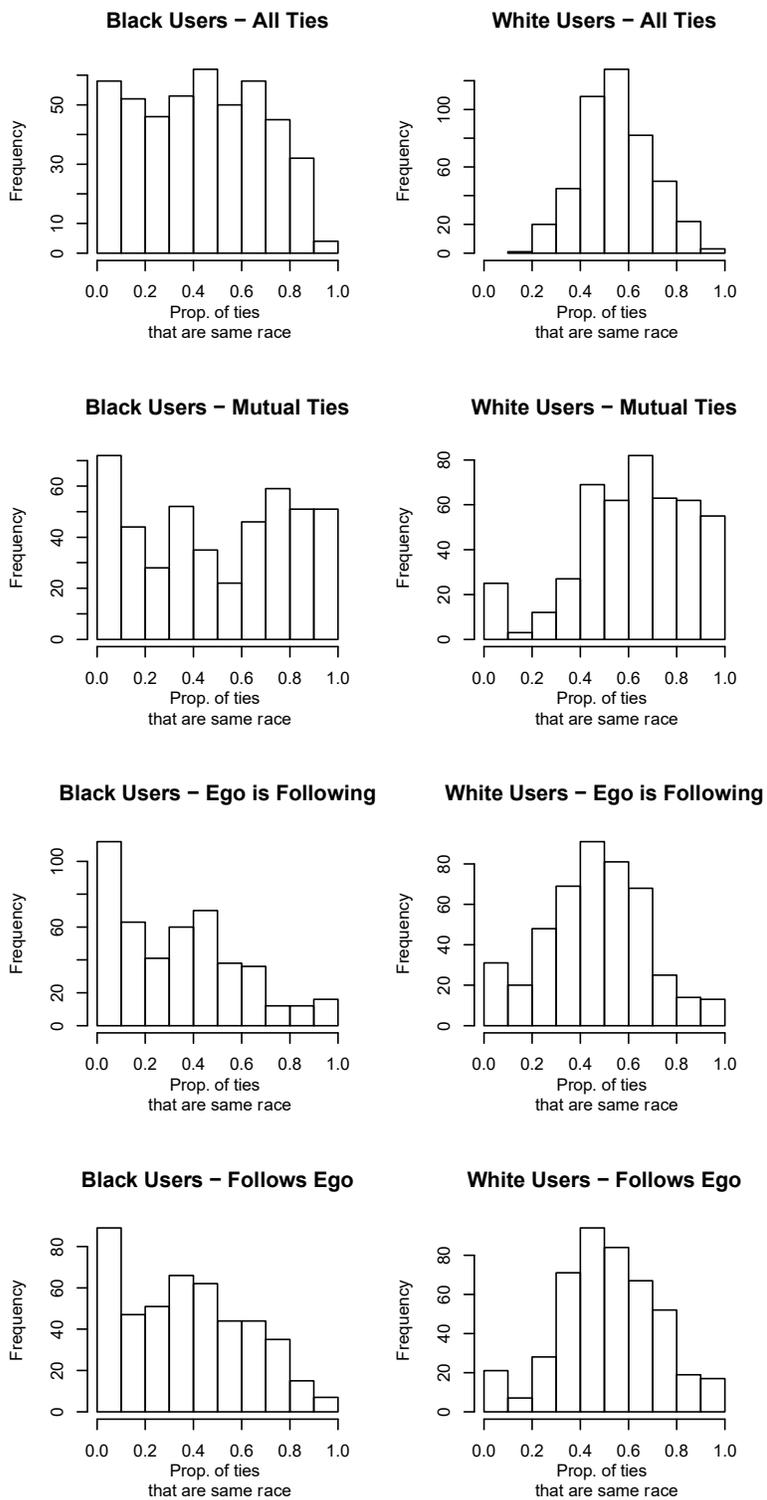

<u>Same race connectedness: Black Egos</u>

| Tie type | Mean | SD | Lower quart | Upper quart | Min | Max |
|---|---|---|---|---|---|---|
| Total ties | 0.423 | 0.246 | 0.206 | 0.626 | 0.000 | 0.923 |
| Mutual | 0.498 | 0.316 | 0.200 | 0.778 | 0.000 | 1.000 |
| Following | 0.340 | 0.268 | 0.106 | 0.500 | 0.000 | 1.000 |
| Follower | 0.358 | 0.247 | 0.145 | 0.550 | 0.000 | 1.000 |

<u>Same race connectedness: White Egos</u>

| Tie type | Mean | SD | Lower Quart | Upper Quart | Min | Max |
|---|---|---|---|---|---|---|
| Total ties | 0.552 | 0.145 | 0.452 | 0.651 | 0.164 | 1.000 |
| Mutual | 0.626 | 0.245 | 0.500 | 0.810 | 0.000 | 1.000 |
| Following | 0.472 | 0.223 | 0.333 | 0.654 | 0.000 | 1.000 |
| Follower | 0.523 | 0.212 | 0.400 | 0.662 | 0.000 | 1.000 |

Mean B/W difference total: $t = -8.8746$, $df = 734.21$, p-value $< 2.2e-16$
Mean B/W difference mutual: $t = -6.6883$, $df = 857.22$, p-value $= 4.071e-11$
Mean B/W difference following: $t = -8.0304$, $df = 888.39$, p-value $= 3.073e-15$
Mean B/W difference follower: $t = -9.3263$, $df = 886.13$, p-value $< 2.2e-16$

**Sensitivity Analysis Set 2: Comparison to Random Connections**

*Sensitivity analysis 2A: All 'unknown' photos featuring people are assumed to be white*

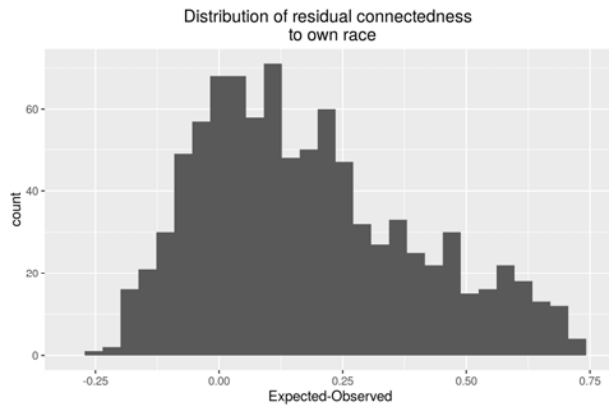

*Figure 1:* Residual Connectedness to Own Race

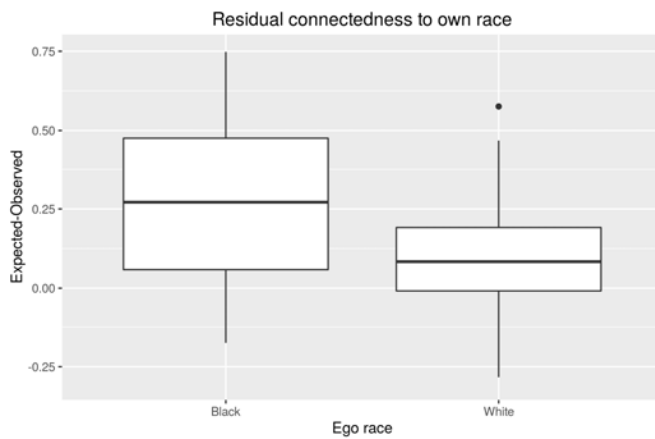

*Figure 2:* Black/White Residual Connectedness to Own Race

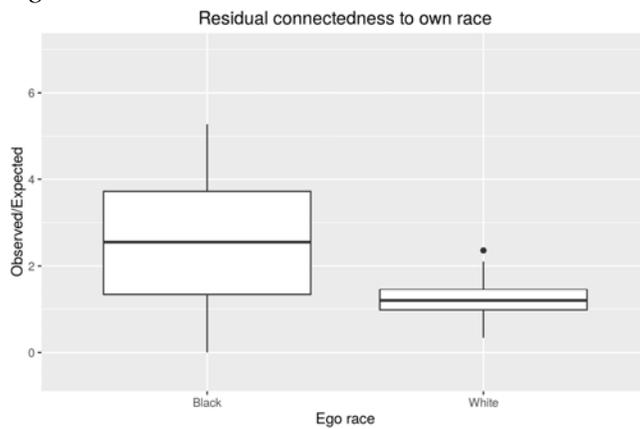

*Figure 3:* Black/White Residual Connectedness to Own Race (Scaled)

*Sensitivity analysis 2B: All person unknowns are white*

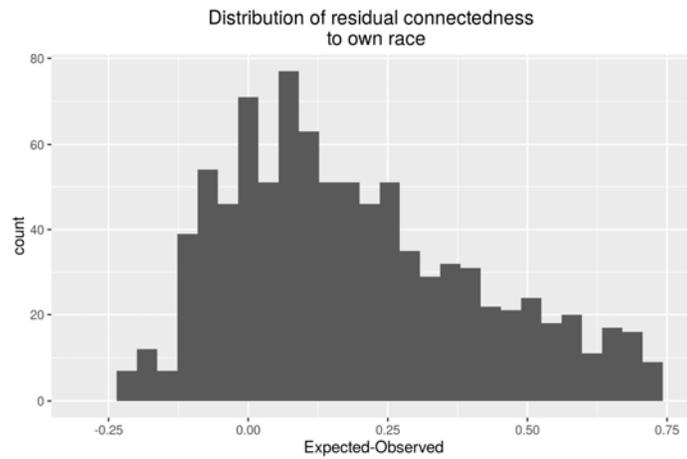

*Figure 1:* Residual Connectedness to Own Race

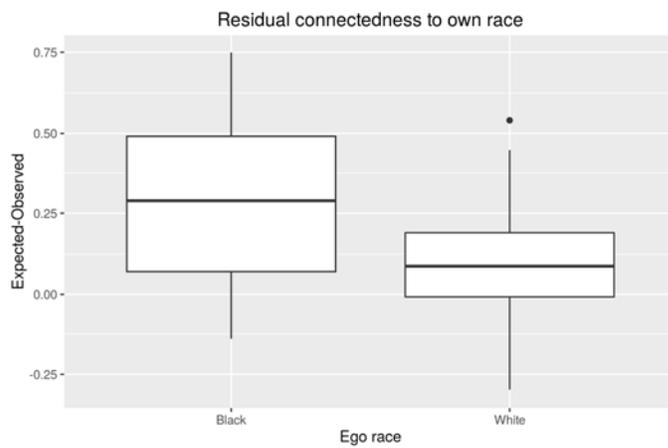

*Figure 2:* Black/White Residual Connectedness to Own Race

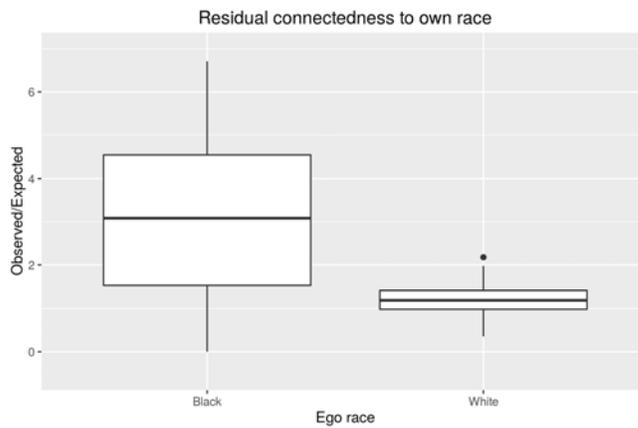

*Figure 3:* Black/White Residual Connectedness to Own Race (Scaled)